\documentclass[preprint,amssymb,amsmath,floatfix]{revtex4}

\pdfoutput=1
\usepackage{t1enc}
\usepackage[latin1]{inputenc}
\usepackage[english]{babel}
\usepackage[dvips]{color}
\usepackage{bm}
\usepackage{latexsym}
\usepackage{longtable}
\usepackage{floatflt}
\usepackage{graphicx}

\begin{document}

\author{Elena Santopinto}
\email{elena.santopinto@ge.infn.it}
\affiliation{INFN and Universit\`a di Genova,
via Dodecaneso 33, 16146 Genova, Italy}

\author{Giuseppe Galat\`a}
\affiliation{INFN and Universit\`a di Genova,
via Dodecaneso 33, 16146 Genova, Italy}
\title{A quark-diquark baryon model}

\begin{abstract}
A simple quark-diquark model for the baryons is constructed as a partial solution to the well known missing resonances problem. A complete classification of the baryonic states in the quark-diquark framework is given and the spectrum is calculated through a mass formula built to reproduce the rotational and vibrational Regge trajectories. 
\end{abstract}

\maketitle

\section{\label{sec:introduzione}Introduction.}

From the introduction of the quarks, the baryons have always been thought as made up of three constituent confined quarks. The light baryons, in particular, have been ordered according to the approximate SU(3)$_{f}$ symmetry, which requires that the baryons belong to the multiplets $[1]_{A}\oplus [8]_{M}\oplus [8]_{M}\oplus [10]_{S}$. However, when we consider the spatially excited resonances, many more states are predicted than observed and on the other hand, states with certain quantum numbers appear in the spectrum at excitation energies much lower than predicted \cite{Nakamura:2010zzi}. Considering only the non-strange sector up to an excitation energy of $2.41\; GeV$, in the average about 45 $N$ states are predicted, but only 12 are established (four- or three-star) and 7 are tentative (two- or one-star) \cite{Nakamura:2010zzi}. This is the so-called missing resonances problem. 
One possible solution to this problem is to describe two correlated quarks inside the baryons by means of the diquark effective degree of freedom.
In this case the number of states predicted are considerably fewer. There have been several studies, ranging from one gluon exchange models to lattice QCD calculations, that have investigated the possibility of diquark correlations and found that they are indeed attractive (see for example \cite{Jaffe:2004ph,Wilczek:2004im,Burden:1996nh,Hecht:2002ej,Alexandrou:2006cq}).
In this article we construct all the allowed states in the framework of the constituent quark-diquark model and we try to assign every known light baryons (with masses smaller than $2\;GeV$ circa) to the appropriate multiplet. Thinking the quark-diquark system as a stringlike object analogous to the quark-antiquark mesons \cite{Iachello:1991re,Iachello:1991fj}, we can, moreover, write a simple mass formula, 
%based on the algebraic model $U(4)_{spatial}\otimes SU(3)_{f}\otimes SU(2)_{s}\otimes SU(3)_{c}$, 
constructed with the aim to reproduce both rotational and vibrational Regge trajectories. 
%In the case of a quark and a diquark bound in a baryon, we use the spatial algebra $U(4)$.
%, which is the compact form of the more correct $U(3,1)$.
%  This algebra includes all states corresponding to rotations and vibrations of a string with a quark and a diquark at its ends and was first used to describe the spatially analogous quark-antiquark meson system by Iachello et al. \cite{Iachello:1991re,Iachello:1991fj}. In those works it was suggested that $U(4)$ should be taken as the spectrum generating algebra of geometric excitations of the quark-antiquark string. 
% If the $U(4)$ algebra is then split in the chain $U(4)\supset SO(4)\supset SO(3)\supset SO(2)$, the mass formula for geometric excitations must be constructed in terms of the respective Casimir operators. The Casimir operators of $U(4)$ are not relevant since they contribute a constant term, while the Casimir operators of $SO(4)$ and $SO(3)$ are taken into account. Thus, their eigenvalues are used to reproduce the vibrational and rotational excitation of the system. Since we do not intend to break the rotational $SO(3)$ symmetry, invariants of $SO(2)$ are not introduced.  

\section{\label{sec:modello}A quark-diquark model for baryons.}
In this model we hypothesize that the baryons are a bound state of two elements, a constituent quark and a constituent diquark.
We think the diquark as two correlated quarks with no internal spatial excitations, or at least we hypothesize that their internal spatial excitations will be higher in energy than the scale of masses of the resonances we will consider, i.e. light resonances up to $2\;GeV$ masses. Actually calculations in a simple, Goldstone-theorem-preserving, rainbow-ladder DSE model \cite{Burden:1996nh,Hecht:2002ej} have confirmed that the first spatially excited diquark, the vector diquark, has a mass much higher than the ground states, the scalar and the axial-vector diquarks. Diquarks are made up of two identical fermions and so they have to satisfy the Pauli principle. Since we consider diquarks with no spatial excitations, their colour-spin-flavour wave functions must be antisymmetric. This limits the possible colour-spin-flavour representations to be only 
\begin{subequations}
\begin{eqnarray}
&  \text{colour \; in} ~ [\bar 3]~ \text{(AS),\; spin-flavour\;in} [21]_{sf}~\text{(S)} & \\
&  \text{colour \;in }[6]~\text{(S)} \text{,\; spin-flavour\; in~} [15]_{sf}~ 
\text{(AS)}.  & 
\end{eqnarray}
\end{subequations}
The decomposition of these SU$_{sf}$(6) representations in terms of SU(3)$_{f}\otimes $ SU(2)$_{s}$ is 
 (in the notation $[\text{flavour\;repr.,\;spin}]$)
\begin{subequations}
\begin{eqnarray}
& [21]_{sf}=[\bar 3,0]\oplus [6,1] & \\
& [15]_{sf}=[\bar 3,1]\oplus [6,0]. &
\end{eqnarray} 
\end{subequations}
Since the baryons must be colourless, we can allow only the diquark states in colour 
$[\bar 3]_{c}$: 
\begin{equation}
  |[\bar 3]_{c},[\bar 3]_{f},0>, |[\bar 3]_{c},[6]_{f},1>.
\end{equation}   
The first of the above states is the scalar (or good) diquark, the second is the axial-vector (or bad) diquark.
In the following we will represent scalar diquarks by their costituent quarks
 (denoted by $s$ if strange, $n$ otherwise) in a square bracket, while axial-vector diquarks are in a brace bracket. 
This choice is not casual, because the explicit expression of diquarks is the commutator of the constituent quarks 
for the scalar ones and the anticommutator for the axial-vector ones.

\section{\label{sec:Pauli}Baryons and the Pauli principle.}
The Pauli principle implies that the baryons must be antisymmetric for exchange of each couple of quarks.
First we describe the application of this principle to the baryons in the three quarks model, in order, then, to underline the differencies with the quark-diquark model.

In the three quarks model we can have the spin-flavor states
\begin{equation}
[6]\otimes [6]\otimes [6]=[56]_{S}\oplus [70]_{M}\oplus [70]_{M}\oplus [20]_{A},
\end{equation}
where the subscripts indicate the symmetry of the state. 

Since we have two different relative angular momenta, we can have symmetric, mixed and antisymmetric spatial parts, independently from the spatial model adopted.

In order to obtain an antisymmetric baryon, we have to combine the spin-flavor-spatial part with the antisymmetric color part. Thus, we need a symmetric spin-flavor-spatial part that can be obtained only through the combinations reported in the left side of Table \ref{tab:spinsaporespazio}.
\begin{table}
\begin{center}
\begin{tabular}{cc|cc}
\hline
\multicolumn{2}{c|}{Three quarks baryons} &  \multicolumn{2}{c}{Quark-diquark baryons} \\
spin-flavor & space & spin-flavor & space \\
\hline 
$[56]_{S}$ & $S$ & $[56]_{S}$ & $S$ \\
$[70]_{M}$ & $M$ & $[70]_{M}$ & $M$  \\
$[20]_{A}$ & $A$ &            &      \\
\hline
\end{tabular}
\caption[smallcaption]{Allowed spin-flavor and spatial combinations in the three quarks model (left) and in the quark-diquark model (right).}
\label{tab:spinsaporespazio}
\end{center}
\end{table}

In the quark-diquark model we can have only the spin-flavor states
\begin{equation}
[21]\otimes [6]=[56]_{S}\oplus [70]_{M}.
\end{equation}

Since in the quark-diquark model we freeze one spatial degree of freedom, thus fixing one of the two relative angular momenta to zero and letting the other vary, we can have only symmetric (if the relative orbital angular momentum $L$ is even) or mixed (if $L$ is odd) spatial parts. We report in the right side of Table \ref{tab:spinsaporespazio} the allowed spin-flavor-space combinations.
% \begin{table}
% \begin{center}
% \begin{tabular}{cc}
% \hline
% spin-flavor & space \\
% \hline 
% $[56]_{S}$ & $S$ \\
% $[70]_{M}$ & $M$ \\
% \hline
% \end{tabular}
% \caption[smallcaption]{Allowed spin-flavor and spatial combinations in the quark-diquark model.}
% \label{tab:qdiqspinsaporespazio}
% \end{center}
% \end{table} 
Hence the sequence of states would be
\begin{equation}
(SU(6)_{sf},L^{P})=([56],0^{+}),([70],1^{-}),([56],2^{+}),
\end{equation}
and so on... .

\section{\label{sec:formulamassa}The mass formula}
% We address the problem to describe the spectrum of baryons with an algebraic model, $U(4)_{spatial}\otimes SU(3)_{f}\otimes SU(2)_{s}\otimes SU(3)_{c}$ \cite{Iachello:1991re,Iachello:1991fj}, based on string-like configurations of a quark-diquark system.
% following what already done for $qqq$ baryons by Bijker, Iachello and Leviatan in Refs. \cite{Bijker:1994yr, Bijker:1996tr, Bijker:2000gq}.

We write for the baryons in the quark-diquark model a simple mass formula which reproduces the Regge trajectories, inspired by the algebraic models for mesons and baryons \cite{Santopinto:2006my,Iachello:1991re,Iachello:1991fj,Bijker:1994yr, Bijker:1996tr, Bijker:2000gq} :
\begin{widetext}
\begin{eqnarray}
\label{eq:formulamassaqdq}
& M^{2}=\Lambda +b\cdot L+c\cdot S(S+1)+d\cdot J+e\cdot I(I+1)+n\cdot \nu +g\cdot C_{2}(SU(3))+h\cdot C_{2}(SU(6))+& \nonumber \\
&+(M_{0}+N_{s}\cdot \Delta M_{s}+N_{[n,s]}\Delta M_{[n,s]}+N_{\{ n,n \}}\cdot \Delta M_{\{ n,n\}}+N_{\{ n,s \}}\cdot \Delta M_{\{ n,s\}}+N_{\{ s,s \}}\cdot \Delta M_{\{ s,s\}})^{2}, &
\end{eqnarray}
\end{widetext}
where $\Lambda $ is an overall scale constant taken equal to $1\;GeV^{2}$, $M_{0}$ the sum of the masses of the non-strange scalar diquark $[n,n]$ and of the non-strange quark, $N_{s}$ and $\Delta M_{s}$ are the number of strange quarks and the mass difference between the strange quark and the non-strange one, $N_{[n,s]}$ and $\Delta M_{[n,s]}$ are the number of strange scalar diquarks and the mass difference between the strange scalar diquark and the non-strange one, $N_{\{ n,s \}}$ and $\Delta M_{\{ n,s \}}$ are the number of strange axial-vector diquarks and the mass difference between the strange axial-vector diquark and the non-strange scalar diquark, $N_{\{ s,s \}}$ and $\Delta M_{\{ s,s\}}$ are the number of double strange axial-vector diquarks and the mass difference between the double strange axial-vector diquark and the non-strange scalar diquark, $C_{2}(SU(3)_{f})$ and $C_{2}(SU(6)_{sf})$ are the quadratic Casimirs of flavour SU(3)$_{f}$ and spin-flavour SU(6)$_{sf}$ respectively, $L$ the relative orbital angular momentum, $S$ the total spin, $J$ the total angular momentum and $\nu $ the vibrational quantum number. 

\begin{figure*}[p]
\includegraphics[width=10cm]{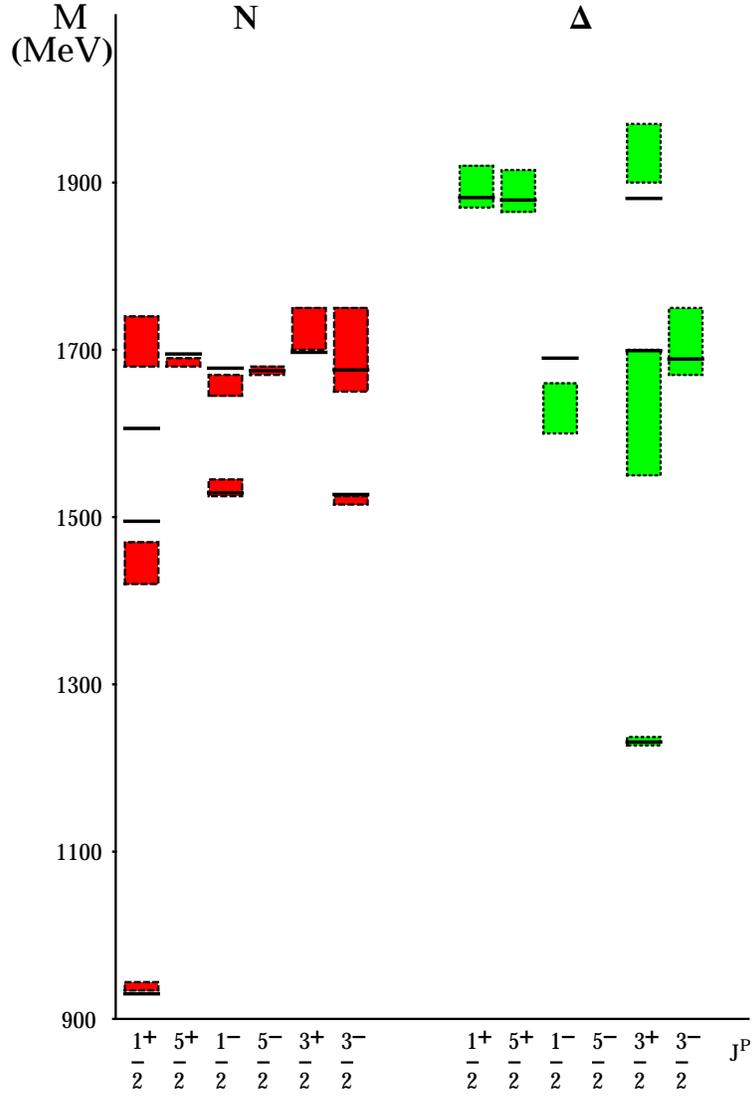}  
\caption{\label{fig:grafnonstrani}(COLOR ONLINE) Comparison between experimental data and fitted results for non-strange baryon resonances. The red boxes with dashed edges and the green ones with dotted edges are respectively the nucleons and the deltas, the black lines are the results from the fit of the mass formula Eq. \ref{eq:formulamassaqdq}.}
\end{figure*} 

\begin{figure*}[p]
\includegraphics[width=8cm]{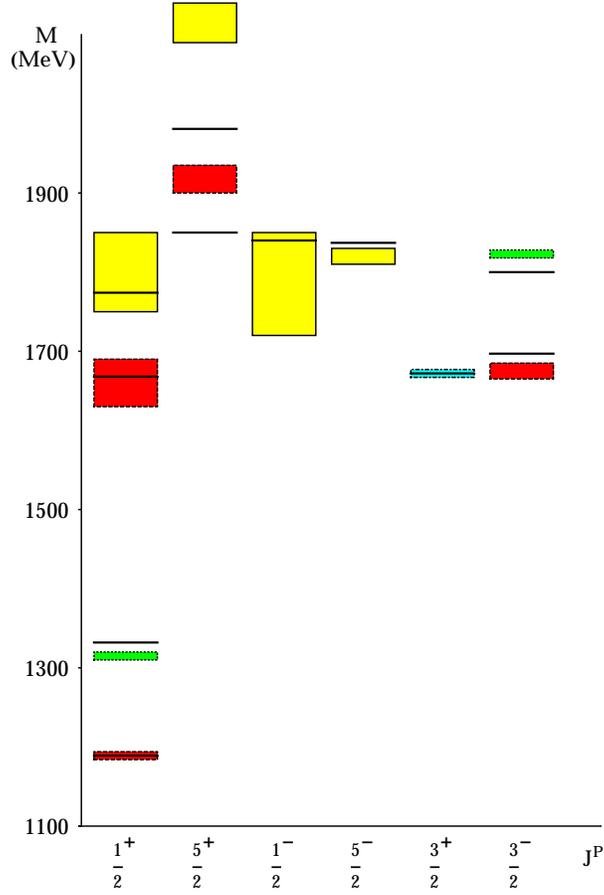}  
\caption{\label{fig:grafstrani}(COLOR ONLINE) Comparison between experimental data and fitted results for strange baryon resonances. The $\Sigma $s, $\Xi $s, $\Omega $ and $\Lambda$s are represented respectively by red boxes with dashed edges, green boxes with dotted edges, light blue boxes with dashdotted edges and yellow boxes with continuous edges, the black lines are the fitted resonances calculated with the mass formula Eq. \ref{eq:formulamassaqdq}.}
\end{figure*}

\section{\label{sec:numeriquantici}Quantum numbers.}
In order to use the mass formula (\ref{eq:formulamassaqdq}), it is necessary to assign to every baryon its quantum numbers, in particular those, like $L$ and $S$, not determined by the experiments. For this purpose we consider only well known baryons, namely the three and four stars baryons.
We classify the light baryons following three guidelines. First of all we must obviously respect the quantum numbers that can be measured experimentally (like $J$, $P$, etc...). Then we must respect the constraint related to the diquark spin-flavor states:
\begin{eqnarray}
 & [21]\otimes [6]=([\bar 3,0]\oplus [6,1])\otimes [3,\frac{1}{2}]= & \nonumber \\
 & =([1,\frac{1}{2}]\oplus [8,\frac{1}{2}])\oplus ([8,\frac{3}{2}]\oplus [8,\frac{1}{2}]\oplus [10,\frac{3}{2}]\oplus [10,\frac{1}{2}]).
\end{eqnarray}
As we can see only the baryons  in a flavor octect and spin $\frac{1}{2}$ can be made up of both the scalar or the vector diquark, while the baryons in a flavor singlet can be composed only by scalar diquarks and those in a flavor decuplet only by vector axial-diquarks.
Finally we must impose that the spin-flavor-space part must be symmetric. As we have seen in section \ref{sec:Pauli}, the consequence is that we must respect the sequence of states $([56],0^{+}),([70],1^{-}),([56],2^{+}),...$, where
\begin{eqnarray}
 & [56]=[10,\frac{3}{2}]\oplus [8,\frac{1}{2}] & \nonumber \\
 & [70]=[10,\frac{1}{2}]\oplus [8,\frac{1}{2}]\oplus [8,\frac{3}{2}]\oplus [1,\frac{1}{2}]. & \nonumber 
\end{eqnarray}
This means, for example, that we cannot have a flavor singlet with $L=0$.

\begin{table}
\begin{center}
\caption{\label{Classgenerale}General classification of the baryon multiplets in the quark-diquark model. $m$ is an integer $\geq 0$, $S_{D}$ is the diquark spin (0 is the scalar diquark, 1 the axial-vector diquark). For $J=\frac{1}{2}$ the states $[8,\frac{3}{2}]$ with $L^{P}=(2m-1)^{-}$ and $[10,\frac{3}{2}]$ with $L^{P}=(2m)^{+}$ are not allowed. The energy splittings and the actual ordering of the various multiplets will obviously depend on the details of the particular model used.}
\begin{tabular}{|c|c|c|c|}
\hline
 $J$ & $L^{P}$ & $S_{D}$ & multiplets ($[SU(3)_{f},Spin]$) \\
\hline
$2m+\frac{1}{2}$ & $(2m)^{+}$ & 0 & $[8,\frac{1}{2}]$ \\
 & $(2m+1)^{-}$ & 0 & $[8,\frac{1}{2}]$,$[1,\frac{1}{2}]$ \\
 & $(2m)^{+}$ & 1 & $[8,\frac{1}{2}]$ \\
 & $(2m+1)^{-}$ & 1 & $[8,\frac{1}{2}]$,$[10,\frac{1}{2}]$ \\
 & $(2m-1)^{-}$ & 1 & $[8,\frac{3}{2}]$ \\
 & $(2m)^{+}$ & 1 & $[10,\frac{3}{2}]$ \\
 & $(2m+1)^{-}$ & 1 & $[8,\frac{3}{2}]$ \\
 & $(2m+2)^{+}$ & 1 & $[10,\frac{3}{2}]$ \\
\hline
$2m+\frac{3}{2}$ & $(2m+1)^{-}$ & 0 & $[8,\frac{1}{2}]$,$[1,\frac{1}{2}]$ \\
 & $(2m+2)^{+}$ & 0 & $[8,\frac{1}{2}]$ \\
 & $(2m+1)^{-}$ & 1 & $[8,\frac{1}{2}]$,$[10,\frac{1}{2}]$ \\ 
 & $(2m+2)^{+}$ & 1 & $[8,\frac{1}{2}]$ \\ 
 & $(2m)^{+}$ & 1 & $[10,\frac{3}{2}]$ \\ 
 & $(2m+1)^{-}$ & 1 & $[8,\frac{3}{2}]$ \\ 
 & $(2m+2)^{+}$ & 1 & $[10,\frac{3}{2}]$ \\ 
 & $(2m+3)^{-}$ & 1 & $[8,\frac{3}{2}]$ \\ 
\hline 
\end{tabular}
\end{center}
\end{table}

\begin{table*}
\begin{center}
\caption{\label{tab:multiplettibarionicidiq} Quark-diquark model assignments for some of the known baryons. Assignments for several states are merely educated guesses. Resonances marked with "no'' are forbidden in the quark-diquark scheme by the rules summed up in section \ref{sec:numeriquantici}. Resonances labelled with "missing'' are the so-called missing resonances. While the missing resonances are still numerous in the quark-diquark scheme, we must underline that they are much less than in the three quark model. In fact in this model most of the "no'' labels would be "missing'' labels.}
\begin{tabular}{|c||c|c|c|c||c|c|c|c||c|}
\hline
 & \multicolumn{4}{||c||}{octet } &  \multicolumn{4}{||c||}{decuplet} & singlet \\
\hline
$J^{P},L,S,S_{D}$ & $\frac{1}{2}$ & $1$ & $0$ & $\frac{1}{2}$ &  $\frac{3}{2}$ & $1$ & $\frac{1}{2}$ & $0$ & $0$\\
\hline
\hline
$\frac{1}{2}^{+},0,\frac{1}{2},0$ &  $N(939)$ & $\Sigma(1189)$ & $\Lambda (1116)$ & $\Xi (1318)$ & no & no & no & no & no \\   
\hline
$\frac{1}{2}^{+},0,\frac{1}{2},1$ & missing & missing & missing & missing &  no & no & no & no & no \\
\hline
$\frac{1}{2}^{+},2,\frac{3}{2},1$ & no & no & no & no &  $\Delta (1910)$ & missing & missing & missing & no \\
\hline
\hline
$\frac{1}{2}^{-},1,\frac{1}{2},0$ &  $N(1535)$ & $\Sigma(1620)$ & $\Lambda (1670)$ & missing & no & no & no & no & $\Lambda (1405)$ \\ 
\hline
$\frac{1}{2}^{-},1,\frac{1}{2},1$ & missing & missing & missing & missing &  $\Delta (1620)$ & missing & missing & missing & no \\
\hline
$\frac{1}{2}^{-},1,\frac{3}{2},1$ & $N(1650)$ & missing & $\Lambda (1800)$ & missing &  no & no & no & no & no \\
\hline
\hline
$\frac{3}{2}^{+},2,\frac{1}{2},0$ &  $N(1720)$ & missing & $\Lambda (1890)$ & missing & no & no & no & no & no \\
\hline
$\frac{3}{2}^{+},0,\frac{3}{2},1$ & no & no & no & no &  $\Delta (1232)$ & $\Sigma(1385)$ & $\Xi (1530)$ & $\Omega (1672)$ & no \\
\hline
$\frac{3}{2}^{+},2,\frac{1}{2},1$ & missing & missing & missing & missing & no & no & no & no & no \\
\hline
$\frac{3}{2}^{+},2,\frac{3}{2},1$ & no & no & no & no &  $\Delta (1920)$ & missing & missing & missing & no \\
\hline
\hline
$\frac{3}{2}^{-},1,\frac{1}{2},0$ &  $N(1520)$ & $\Sigma(1670)$ & $\Lambda (1690)$ & $\Xi (1820)$ & no & no & no & no & $\Lambda (1520)$ \\
\hline
$\frac{3}{2}^{-},1,\frac{1}{2},1$ & missing & missing & missing & missing &  $\Delta (1700)$ & missing & missing & missing & no \\
\hline
$\frac{3}{2}^{-},1,\frac{3}{2},1$ & $N(1700)$ & $\Sigma(1940)$ & $\Lambda (1960)$ & missing &  no & no & no & no & no \\
\hline
$\frac{3}{2}^{-},3,\frac{3}{2},1$ & missing & missing & missing & missing &  no & no & no & no & no \\
\hline
\hline
$\frac{5}{2}^{+},2,\frac{1}{2},0$ &  $N(1680)$ & $\Sigma(1915)$ & $\Lambda (1820)$ & $\Xi (2030)$ & no & no & no & no & no \\
\hline
$\frac{5}{2}^{+},2,\frac{1}{2},1$ & missing & missing & $\Lambda (2110)$ & missing & no & no & no & no & no \\
\hline
$\frac{5}{2}^{+},2,\frac{3}{2},1$ & no & no & no & no &  $\Delta (1905)$ & missing & missing & missing & no \\
\hline
$\frac{5}{2}^{+},4,\frac{3}{2},1$ & no & no & no & no & missing & missing & missing & missing & no \\
\hline
\hline
$\frac{5}{2}^{-},3,\frac{1}{2},0$ &  missing & missing & missing & missing & no & no & no & no & missing \\
\hline
$\frac{5}{2}^{-},3,\frac{1}{2},1$ & missing & missing & missing & missing & $\Delta (1930)$ & missing & missing & missing & no \\
\hline
$\frac{5}{2}^{-},1,\frac{3}{2},1$ & $N(1675)$ & $\Sigma(1775)$ & $\Lambda (1830)$ & missing &  no & no & no & no & no \\
\hline
$\frac{5}{2}^{-},3,\frac{3}{2},1$ & missing & missing & missing & missing &  no & no & no & no & no \\
\hline
\end{tabular}
\end{center}
\end{table*}
In Table \ref{Classgenerale} we report a general classification, valid for all quantum numbers, of the baryon multiplets in the quark-diquark model, while in Table \ref{tab:multiplettibarionicidiq} we assign the light known baryons to each multiplet. The missing and the not allowed states are reported in the table. These tables are in part based on the analogous tables compiled by Bijker, Iachello and Leviatan \cite{Bijker:2000gq}, Selem and Wilczek \cite{Selem:2006nd} and the PDG \cite{Nakamura:2010zzi}. It is important to underline that we lack a sure criterion to assign the diquark content to the baryons (i.e. we cannot say if a particular baryon should be made up of a scalar, an axial-vector or even a mixing of the two diquarks). We found only two sure elements on which the choice can be based:
\begin{itemize}
\item Isospin and strangeness:

We must remember that every baryon family has a definite isospin and strangeness: $N$ has isospin $I=\frac{1}{2}$ and strangeness $\mathcal{S}=0$, $\Delta $ has $I=\frac{3}{2}$ and $\mathcal{S}=0$, $\Lambda $ has $I=0$ and $\mathcal{S}=-1$, $\Sigma $ has $I=1$ and $\mathcal{S}=-1$, $\Xi $ has $I=\frac{1}{2}$ and $\mathcal{S}=-2$, $\Omega $ has $I=0$ and $\mathcal{S}=-3$. Thus, we must combine the diquark and the quark to reproduce isospin and strangeness of the baryon.
But we can easily find that $[n,n]$ has $I=0$ and $\mathcal{S}=0$, $[n,s]$ has $I=\frac{1}{2}$ and $\mathcal{S}=-1$, $\{ n,n\}$ has $I=1$ and $\mathcal{S}=0$, $\{ n,s\}$ has $I=\frac{1}{2}$ and $\mathcal{S}=-1$ and $\{ s,s\}$ has $I=0$ and $\mathcal{S}=-2$. Combining the quark and the diquark together we find the possible diquark content.
$N$ can be either $[n,n]n$ or $\{ n,n\}n$, $\Delta $ can be only $\{ n,n\}n$, $\Lambda $ can be $[n,n]s$ or $[n,s]n$ if they are in a flavour singlet otherwise they can be $[n,n]s$, $[n,s]n$ or $\{ n,s\}n$ if they belong to a flavour octect, $\Sigma $ can be $[n,s]n$, $\{ n,n\} s$ or $\{ n,s\}n$, $\Xi $ can be $[n,s]s$, $\{ n,s\} s$ or $\{ s,s\}n$ and finally $\Omega $ can be only $\{ s,s\}s$.
\item Diquark masses:

We can say, following all the previous studies about the diquarks (as for example Refs. \cite{Jaffe:2004ph,Wilczek:2004im,Alexandrou:2006cq}), that the axial-vector diquark should be heavier than the scalar one. Thus, if we have two baryons with similar quantum numbers but different masses, we will assign the axial-vector diquark to the heavier one.
\end{itemize}
In this first attempt we choose to assign to all the baryons being part of the same flavour multiplet an analogous diquark content(i.e. if we establish that a baryon should have for example a scalar diquark, then all the other baryons of the same multiplet will have a scalar diquark). In this way we think that all the mass differences inside a baryon multiplet should be addressed to the different strangeness of the various baryons.

\section{\label{sec:risultati}Fit and results}

\begin{table*}
\begin{center}
\caption{\label{tab:risonanzebarionidiqL0} Quantum numbers of the baryonic resonances with orbital angular momentum $L=0$ in the quark-diquark model. In the section (a) we list the resonances used for the fit with our mass formula Eq. \ref{eq:formulamassaqdq}, while in section (b) there are the resonances we could not use because we were not sure about their diquark content, in this case the diquark content is marked with a question mark, or simply because they are not well established. All the masses are in $GeV$.}
\begin{tabular}{|c|c|c|c|c|c|c|c|c|}
\hline
Resonances & $J$ & $S$ & composition & $SU(3)_{f}$ multiplet & $SU(6)_{sf}$ multiplet & $\nu $ & $M(exp)$ & $M(theo)$ \\
\hline
 \multicolumn{8}{|c|}{(a)} \\
\hline
$N(939)$ & $\frac{1}{2}$ & $\frac{1}{2}$ & $[n,n]n$ & $8$ & $56$ & 0 & $0.939\pm 0.005$ & $0.930$ \\
\hline
$\Sigma (1189)$ & $\frac{1}{2}$ & $\frac{1}{2}$ & $[n,s]n$ & $8$ & $56$ & 0 & $1.189\pm 0.005$ & $1.189$ \\
\hline
$\Xi (1318)$ & $\frac{1}{2}$ & $\frac{1}{2}$ & $[n,s]s$ & $8$ & $56$ & 0  & $1.315\pm 0.005$ & $1.332$ \\
\hline
$\Delta (1232)$ & $\frac{3}{2}$ & $\frac{3}{2}$ & $\{n,n\}n$ & $10$ & $56$ & 0 & $1.231-1.233$ & $1.231$ \\
\hline
$\Omega (1672)$ & $\frac{3}{2}$ & $\frac{3}{2}$ & $\{s,s\}s$ & $10$ & $56$ & 0 & $1.672\pm 0.005$ & $1.672$ \\
\hline
$N(1440)$ & $\frac{1}{2}$ & $\frac{1}{2}$ & $[n,n]n$ & $8$ & $56$ & 1 & $1.420-1.470$ & $1.495$ \\
\hline
$\Sigma (1660)$ & $\frac{1}{2}$ & $\frac{1}{2}$ & $[n,s]n$ & $8$ & $56$ & 1 & $1.630-1.690$ & $1.668$ \\
\hline
$N(1710)$ & $\frac{1}{2}$ & $\frac{1}{2}$ & $\{n,n\}n$ & $8$ & $56$ & 1 & $1.680-1.740$ & $1.606$ \\
\hline
$\Lambda (1810)$ & $\frac{1}{2}$ & $\frac{1}{2}$ & $\{n,s\}n$ & $8$ & $56$ & 1 & $1.750-1.850$ & $1.774$ \\
\hline
$\Delta (1600)$ & $\frac{3}{2}$ & $\frac{3}{2}$ & $\{n,n\}n$ & $10$ & $56$ & 1 & $1.550-1.700$ & $1.699$ \\
\hline
 \multicolumn{8}{|c|}{(b)} \\
\hline
$\Lambda (1116)$ & $\frac{1}{2}$ & $\frac{1}{2}$ & $[n,n]s\;(?)$ & $8$ & $56$ & 0 & $1.116\pm 0.005$ & $1.087$  \\
\hline
$\Sigma (1385)$ & $\frac{3}{2}$ & $\frac{3}{2}$ & $\{n,n\}s\;(?)$ & $10$ & $56$ & 0 & $1.383\pm 0.005$ & $1.359$ \\
\hline
$\Xi (1530)$ & $\frac{3}{2}$ & $\frac{3}{2}$ & $\{s,s\}n\;(?)$ & $10$ & $56$ & 0 & $1.532\pm 0.005$ & $1.537$\\
\hline
$\Lambda (1600)$ & $\frac{1}{2}$ & $\frac{1}{2}$ & $[n,n]s\; (?)$ & $8$ & $56$ & 1 & $1.560-1.700$ & $1.598$ \\
\hline
\end{tabular}
\end{center}
\end{table*}

\begin{table*}
\begin{center}
\caption{\label{tab:risonanzebarionidiqL1} Same as Table \ref{tab:risonanzebarionidiqL0} but for baryons with $L=1$. All the masses are in $GeV$.}
\begin{tabular}{|c|c|c|c|c|c|c|c|c|}
\hline
Resonances & $J$ & $S$ & composition & $SU(3)_{f}$ multiplet & $SU(6)_{sf}$ multiplet & $\nu $ & $M(exp)$ & $M(theo)$ \\
\hline
 \multicolumn{8}{|c|}{(a)} \\
\hline
$N(1535)$ & $\frac{1}{2}$ & $\frac{1}{2}$ & $[n,n]n$ & $8$ & $70$ & 0 & $1.525-1.545$ & $1.529$ \\
\hline 
$N(1520)$ & $\frac{3}{2}$ & $\frac{1}{2}$ & $[n,n]n$ & $8$ & $70$ & 0 & $1.515-1.525$ & $1.527$ \\
\hline
$\Sigma (1670)$ & $\frac{3}{2}$ & $\frac{1}{2}$ & $[n,s]n$ & $8$ & $70$ & 0 & $1.665-1.685$ & $1.697$ \\
\hline
$\Xi (1820)$ & $\frac{3}{2}$ & $\frac{1}{2}$ & $[n,s]s$ & $8$ & $70$ & 0 & $1.818-1.828$ & $1.800$ \\
\hline
$N(1650)$ & $\frac{1}{2}$ & $\frac{3}{2}$ & $\{n,n\}n$ & $8$ & $70$ & 0 & $1.645-1.670$ & $1.678$ \\
\hline
$\Lambda (1800)$ & $\frac{1}{2}$ & $\frac{3}{2}$ & $\{n,s\}n$ & $8$ & $70$ & 0 & $1.720-1.850$ & $1.840$ \\
\hline
$N(1700)$ & $\frac{3}{2}$ & $\frac{3}{2}$ & $\{n,n\}n$ & $8$ & $70$ & 0 & $1.650-1.750$ & $1.676$ \\
\hline   
$N(1675)$ & $\frac{5}{2}$ & $\frac{3}{2}$ & $\{n,n\}n$ & $8$ & $70$ & 0 & $1.670-1.680$ & $1.675$ \\
\hline
$\Lambda (1830)$ & $\frac{5}{2}$ & $\frac{3}{2}$ & $\{n,s\}n$ & $8$ & $70$ & 0 & $1.810-1.830$ & $1.837$ \\
\hline
$\Delta (1620)$ & $\frac{1}{2}$ & $\frac{1}{2}$ & $\{n,n\}n$ & $10$ & $70$ & 0 & $1.600-1.660$ & $1.690$ \\
\hline
$\Delta (1700)$ & $\frac{3}{2}$ & $\frac{1}{2}$ & $\{n,n\}n$ & $10$ & $70$ & 0 & $1.670-1.750$ & $1.689$ \\
\hline
 \multicolumn{8}{|c|}{(b)} \\
\hline
$\Lambda (1405)$ & $\frac{1}{2}$ & $\frac{1}{2}$ & $[n,n]s\;(?)$ & $8$ & $70$ & 0 & $1.402-1.410$ & $1.593$ \\
\hline
$\Lambda (1520)$ & $\frac{3}{2}$ & $\frac{1}{2}$ & $[n,n]s\;(?)$ & $8$ & $70$ & 0 & $1.520\pm 0.005$ & $1.591$  \\
\hline  
$\Lambda (1670)$ & $\frac{1}{2}$ & $\frac{1}{2}$ & $[n,s]n\;(?)$ & $8$ & $70$ & 0 & $1.660-1.680$ & $1.687$ \\
\hline 
$\Lambda (1690)$ & $\frac{3}{2}$ & $\frac{1}{2}$ & $[n,s]n\;(?)$ & $8$ & $70$ & 0 & $1.685-1.695$ & $1.685$ \\
\hline
$\Sigma (1750)$ & $\frac{1}{2}$ & $\frac{3}{2}$ & $\{n,n\}s\;(?)$ & $8$ & $70$ & 0 & $1.730-1.800$ & $1.753$\\
\hline
$\Lambda (1960)$ & $\frac{3}{2}$ & $\frac{3}{2}$ & $\{n,s\}n$ & $8$ & $70$ & 0 &  & $1.839$ \\
\hline
$\Sigma (1940)$ & $\frac{3}{2}$ & $\frac{3}{2}$ & $\{n,n\}s\;(?)$ & $8$ & $70$ & 0 & $1.900-1.950$ & $1.850$\\
\hline
$\Sigma (1775)$ & $\frac{5}{2}$ & $\frac{3}{2}$ & $\{n,n\}s\;(?)$ & $8$ & $70$ & 0 & $1.770-1.780$ & $1.788$\\
\hline
\end{tabular}
\end{center}
\end{table*}

\begin{table*}
\begin{center}
\caption{\label{tab:risonanzebarionidiqL2} Same as Table \ref{tab:risonanzebarionidiqL0} but for baryons with $L=2$. All the masses are in $GeV$.}
\begin{tabular}{|c|c|c|c|c|c|c|c|c|}
\hline
Resonances & $J$ & $S$ & composition & $SU(3)_{f}$ multiplet & $SU(6)_{sf}$ multiplet & $\nu $ & $M(exp)$ & $M(theo)$ \\
\hline
 \multicolumn{8}{|c|}{(a)} \\
\hline
$N(1720)$ & $\frac{3}{2}$ & $\frac{1}{2}$ & $[n,n]n$ & $8$ & $56$ & 0 & $1.700-1.750$ & $1.697$ \\
\hline
$N(1680)$ & $\frac{5}{2}$ & $\frac{1}{2}$ & $[n,n]n$ & $8$ & $56$ & 0 & $1.680-1.690$ & $1.695$ \\
\hline
$\Sigma (1915)$ & $\frac{5}{2}$ & $\frac{1}{2}$ & $[n,s]n$ & $8$ & $56$ & 0 & $1.900-1.935$ & $1.850$ \\
\hline
$\Lambda (2110)$ & $\frac{5}{2}$ & $\frac{3}{2}$ & $\{n,s\}n$ & $8$ & $56$ & 0 & $2.090-2.140$ & $1.981$ \\
\hline
$\Delta (1910)$ & $\frac{1}{2}$ & $\frac{3}{2}$ & $\{n,n\}n$ & $10$ & $56$ & 0 & $1.870-1.920$ & $1.882$ \\
\hline
$\Delta (1920)$ & $\frac{3}{2}$ & $\frac{3}{2}$ & $\{n,n\}n$ & $10$ & $56$ & 0 & $1.900-1.970$ & $1.881$ \\
\hline
$\Delta (1905)$ & $\frac{5}{2}$ & $\frac{3}{2}$ & $\{n,n\}n$ & $10$ & $56$ & 0 & $1.865-1.915$ & $1.879$ \\
\hline
$\Delta (1950)$ & $\frac{7}{2}$ & $\frac{3}{2}$ & $\{n,n\}n$ & $10$ & $56$ & 0 & $1.915-1.950$ & $1.878$ \\
\hline
 \multicolumn{8}{|c|}{(b)} \\
\hline
$\Lambda (1890)$ & $\frac{3}{2}$ & $\frac{1}{2}$ & $[n,s]n\;(?)$ & $8$ & $56$ & 0 & $1.850-1.910$ & $1.841$ \\
\hline
$\Lambda (1820)$ & $\frac{5}{2}$ & $\frac{1}{2}$ & $[n,s]n\;(?)$ & $8$ & $56$ & 0 & $1.815-1.825$ & $1.839$ \\
\hline
$\Sigma (1880)$ & $\frac{1}{2}$ & $\frac{3}{2}$ & $\{n,n\}s\;(?)$ & $8$ & $56$ & 0 & $1.880$ & $1.939$ \\
\hline 
$N(2000)$ & $\frac{5}{2}$ & $\frac{3}{2}$ & $\{n,n\}n$ & $8$ & $56$ & 0 &  & $1.831$ \\
\hline
$\Sigma (2080)$ & $\frac{3}{2}$ & $\frac{3}{2}$ & $\{n,s\}n\;(?)$ & $10$ & $56$ & 0 & $2.080$ & $2.022$ \\
\hline
$\Sigma (2070)$ & $\frac{5}{2}$ & $\frac{3}{2}$ & $\{n,s\}n\;(?)$ & $10$ & $56$ & 0 & $2.070$ & $2.020$ \\
\hline
$\Sigma (2030)$ & $\frac{7}{2}$ & $\frac{3}{2}$ & $\{n,s\}n\;(?)$ & $10$ & $56$ & 0 & $2.025-2.040$ & $2.019$\\
\hline
\end{tabular}
\end{center}
\end{table*}

We determine now the parameters of the mass formula (\ref{eq:formulamassaqdq}) through a fit. We excluded from the fit  the states for which their diquark content cannot be determined following the criterion described in section \ref{sec:numeriquantici}. These states have a question mark next to their diquark content in Tables \ref{tab:risonanzebarionidiqL0}, \ref{tab:risonanzebarionidiqL1} and \ref{tab:risonanzebarionidiqL2}.
The results of the fit are:
\begin{subequations} 
\begin{eqnarray}
 M_{0} & =  & (1.197\pm 0.015) \; GeV\\
 \Delta M_{s}         & =  & (0.132\pm 0.007) \; GeV\\
 \Delta M_{[n,s]}     & =  & (0.201\pm 0.005) \; GeV\\
 \Delta M_{\{ n,n\}}  & =  & (0.135\pm 0.024) \; GeV\\
 \Delta M_{\{ n,s\}}  & =  & (0.339\pm 0.023) \; GeV\\
 \Delta M_{\{ s,s\}}  & =  & (0.441\pm 0.019) \; GeV\\
 b      & =  & (1.011\pm 0.016) \; GeV^2\\
 c      & =  & (0.046\pm 0.022) \; GeV^2\\
 d      & =  & (-0.006\pm 0.015) \; GeV^2\\
 e      & =  & (0.020\pm 0.008) \; GeV^2\\
 n      & =  & (1.37\pm 0.05) \; GeV^2\\
 g      & =  & (0.039\pm 0.007) \; GeV^2\\
 h      & =  & (-0.154\pm 0.004) \; GeV^2
\end{eqnarray}
\end{subequations}

\section{\label{sec:discussione}Critical discussion of the results.}
The principal feature of our quark-diquark model is the drastic cut in the number of baryonic states. In fact, while all the existing baryonic resonances still fits well in our scheme, we have much less missing states than a normal three quarks constituent model. Nevertheless quite a few missing states still remain and these should be further investigated, both from a theoretical and an experimental point of view. 
\begin{table*}
\begin{center}
\caption{\label{tab:massediquarkaltri} Mass differences (in $GeV$) between scalar and axial-vector diquarks according to some important studies, compared with the results obtained in this work.}
\begin{tabular}{|c|c|c|c|c|c|c|}
\hline
$M_{[n,n]}$ & $M_{\{ n,n \} }-M_{[n,n]}$ & $M_{[n,s]}-M_{[n,n]}$ & $M_{\{ n,s \} }-M_{[n,s]}$ & $M_{\{ n,s \} }-M_{\{ n,n \} }$ & $M_{\{ s,s \} }-M_{\{ n,s \} }$ & Source \\
\hline
0.688 & 0.202 & 0.272 & - & - & - & Maris \cite{Maris:2002yu,Maris:2004ig} \\
\hline
- & 0.29 & - & 0.11 & - & - & Wilczek \cite{Wilczek:2004im} \\
\hline
- & 0.210 & - & 0.150 & - & - & Jaffe \cite{Jaffe:2004ph} \\
\hline
0.595 & 0.205 & 0.240 & 0.140 & 0.175 & - & Lichtenberg \cite{Lichtenberg:1996fi} \\
\hline
0.74 & 0.21 & 0.14 & 0.17 & 0.10 & 0.08 & Roberts \cite{Burden:1996nh,Hecht:2002ej} \\
\hline
- & 0.135 & 0.201 & 0.138 & 0.204 & 0.101 & This work \\
\hline
\end{tabular}
\end{center}
\end{table*}
The mass formula resulting from the fit describes reasonably well the spectrum, with a $\chi ^{2}/n.d.f=8.75$. The resulting orbital and vibrational Regge trajectory slopes, $\alpha =b+d=1.005\;GeV^{2}$ and $n=1.37\;GeV^{2}$, agree quite well with the theoretical expectations in a string model \cite{Santopinto:2006my,'tHooft:1974hx,Johnson:1975sg,Iachello:1991re,Iachello:1991fj}. Important parameters of constituent quark models are, more than the absolute masses of the constituent quarks, which can vary greatly with the model used, the mass differences between these constituents, which tend to be more stable and may be compared with results obtained with both constituent and other models, such as QCD inspired and lattice ones. Our value for the mass difference between the strange and the non-strange quark $\Delta M_{s}$ is compatible with the estimates of the constituent quark models and with the PDG value for the current quark mass difference \cite{Nakamura:2010zzi}. The difference $\Delta M_{\{ n,n\}}=M_{\{ n,n \} }-M_{[n,n]}$, as well as the mass differences between $[n,s]$ and $[n,n]$, between $\{ n,s\}$ and $\{ n,n\}$ and between $\{ n,s \}$ and $[n,s]$, have been compared with the predictions made through the main other models for the constituent diquark (see Table \ref{tab:massediquarkaltri}). Apart from $\Delta M_{\{ n,n\}}$, which is somewhat smaller than the other models, all the mass differences lie in the same range of values of the other works.

We managed to describe in a sufficiently satisfactory way the baryons spectrum with a very simple mass formula, based essentially on only two elements: the constituent quark-diquark structure of the baryons and the Regge trajectories. Thus, we can conclude that these two elements should be the basis of future, more advanced investigations.

\bibliography{Bibliografia}
\end{document}